\documentclass[a4paper,aps,floats,intlimits,twocolumn,pre]{revtex4-1}
\usepackage{amsmath,amsthm,amssymb,amsthm,dsfont,mathrsfs,bbm}
\usepackage{mathtools}
\usepackage{multirow}
\usepackage{tikz}
\usepackage{tkz-berge,tkz-graph}
\usetikzlibrary{positioning,chains,fit,shapes,calc}
\usepackage{asymptote}
\usepackage[caption=false]{subfig}
\usepackage[english]{babel}
\usepackage{pgfplots}
\pgfplotsset{compat=1.5}

\newcommand{\BM}{\mathbf M}
\newcommand{\Bm}{\mathbf m}
\newcommand{\Bb}{\mathbf{b}}
\newcommand{\Br}{\mathbf{r}}

\newcommand{\mM}{\boldsymbol{\mathcal{M}}}
\newcommand{\mB}{\mathcal{B}}
\newcommand{\mR}{\mathcal{R}}
\newcommand{\mS}{\mathcal{S}}

\newcommand{\R}{\mathds{R}}
\newcommand{\Z}{\mathds{Z}}
\newcommand{\N}{\mathds{N}}
\newcommand{\J}{\mathsf{J}}    
\DeclareMathOperator{\dd}{d}		
\DeclareMathOperator{\e}{e}
\newcommand{\abs}[1]{{\left|#1\right|}}
\newcommand{\norm}[1]{{\left\|#1\right\|}}
\DeclareMathOperator{\Hess}{Hess} 
\newcommand*\pFqskip{8mu}
\catcode`,\active
\newcommand*\pFq{\begingroup
        \catcode`\,\active
        \def ,{\mskip\pFqskip\relax}%
        \dopFq
}
\catcode`\,12
\def\dopFq#1#2#3#4#5{%
        {}_{#1}F_{#2}\biggl[\genfrac..{0pt}{}{#3}{#4};#5\biggr]%
        \endgroup
}
\begin{document}
\title{Scaling hypothesis for the Euclidean bipartite matching problem\\II. Correlation functions}
\author{Sergio Caracciolo}
\affiliation{Dipartimento di Fisica, University of Milan and INFN, via Celoria 16, I-20133 Milan, Italy}
\email{sergio.caracciolo@mi.infn.it}
\author{Gabriele Sicuro}
\affiliation{Centro Brasileiro de Pesquisas F\'isicas, Rua Xavier Sigaud 150, 22290-180 Rio de Janeiro, Brazil}
\email{sicuro@cbpf.br}
\date{\today}
\begin{abstract}
We analyze the random Euclidean bipartite matching problem on the hypertorus in $d$ dimensions with quadratic cost and we derive the two--point correlation function for the optimal matching, using a proper ansatz introduced by \textcite{Caracciolo2014} to evaluate the average optimal matching cost. We consider both the grid--Poisson (\textsc{gP}) matching problem and the Poisson--Poisson (PP) matching problem. We also show that the correlation function is strictly related to the Green's function of the Laplace operator on the analyzed manifold.
\end{abstract}
\maketitle

\section{The matching problem: an introduction}\label{Intro}
The \textit{assignment problem}, or bipartite matching problem, is a classical combinatorial optimization problem in which two sets of $N$ elements, $\mB=\{\Bb_i\}_{i=1,\dots,N}$ and $\mR\coloneqq\{\Br_i\}_{i=1,\dots,N}$ are considered. An assignment is an element $\pi\in\mS_N$ of the set $\mS_N$ of permutations of $N$ elements, such that $\Br_i\mapsto\Bb_{\pi(i)}$. Moreover, a cost function $w\colon\mR\times\mB\to\R^+$, $(\Br_i,\Bb_j)\mapsto w_{ij}$, is given, from which we can define the total cost of a certain assignment $\pi$ as
\begin{equation}
E_N[\pi;w]\coloneqq\frac{1}{N}\sum_{i=1}^Nw_{i\pi(i)}.
\end{equation}
In the bipartite matching problem we want to find the permutation $\pi^*$ that minimizes the previous quantity for a given function $w$. From a computational point of view, the problem belongs to the P computational complexity class and it can be efficiently solved using fast algorithms \cite{Kuhn,Munkres1957,Edmonds1972}. If random instances are considered, i.e. $w_{ij}$ are random quantities, we are usually interested on the average optimal cost,
\begin{equation}
E_N\coloneqq\overline{E_N[\pi^*;w]}
\end{equation}
where we denoted by $\overline{\bullet}$ the expectation over all the possible instances $w$. If the values $\{w_{ij}\}$ are independent and identically distributed random variables, the problem is usually called \textit{random assignment problem}: in this case, the average optimal cost and its properties in the large $N$ limit were investigated both with statistical physics techniques \cite{Mezard1985} and probability arguments \cite{Aldous2001}.

In a more complicated variation of the random assignment problem, the so called \textit{Euclidean bipartite matching problem} (\textsc{Ebmp}), the two sets $\mR$ and $\mB$ are in one-to-one correspondence with uniformly generated random points on the unit hypercube $\Omega_d\coloneqq[0,1]^d\subset\R^d$, whilst the weight $w_{ij}$ is a function of the Euclidean distance $\norm{\Br_i-\Bb_j}$ (for simplicity, we identify the elements $\Bb_i$, $\Br_j$ with the corresponding geometric points in $\Omega_d$): in this case correlations between different values $\{w_{ij}\}$ appear and a proper mathematical treatment is more complicated. In the following we will consider weight functions in the form
\begin{equation}
w_{ij}\coloneqq\norm{\Br_i-\Bb_j}^p,\quad p\in\R^+.
\end{equation}
\textcite{Mezard1988} considered the previous problem on the hypercube $\Omega_d$ for any value of $d$ in the large $N$ limit, assuming that correlations can be treated as perturbations to the purely random case and evaluating approximately the average optimal cost through replica arguments. An exact solution to the problem for the $d=1$ and $p>1$ case is provided in \citep{Boniolo2012,Caracciolo2014b}, where the average optimal cost and correlation functions are computed; moreover, the correspondence between the matching problem and the Brownian bridge process on the line and the circle is proved. 

Denoting by $\pi^*$ the optimal permutation for a given instance, we introduce the \textit{optimal matching ray}:
\begin{equation}
\label{mappa}
\Bm(\Br_i)\coloneqq \Bb_{\pi^*(i)}-\Br_{i},\quad i=1,\dots N.
\end{equation}
The optimal cost is
\begin{align}	&\textstyle
E_N^{(p)}[\Bm;\{\mR,\mB\}]\coloneqq\frac{1}{N}\sum_{i=1}^N\norm{\Bm(\mathbf r_i)}^p,\\&\textstyle E_N^{(p)}(d)\coloneqq \overline{E_N^{(p)}[\Bm;\{\mR,\mB\}]}
\end{align}
where $d$ is the dimensionality of the Euclidean space and the average $\overline{\bullet}$ is performed over the positions of the points. 

\begin{figure}\label{toro}
\includegraphics[width=\columnwidth]{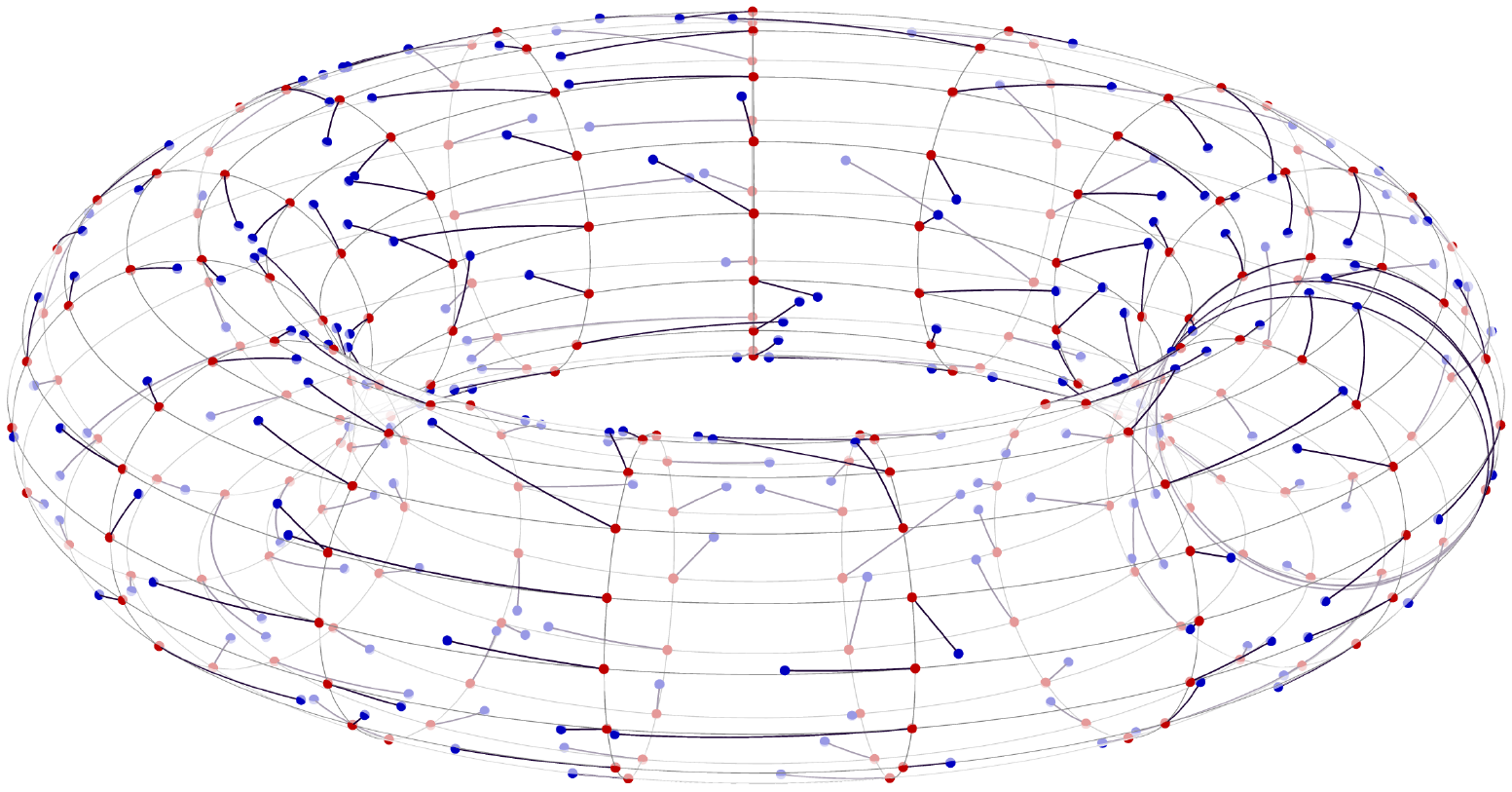}
\caption{\textsc{gP} Euclidean bipartite matching with $N=225$ and $p=2$ on the torus.}
\end{figure}

The scaling properties of the optimal matching ray, and therefore of the optimal cost, are known to the literature for $p>1$ \cite{Caracciolo2014b,Ajtai1984,Talagrand1992}, being
\begin{equation}
\norm{\Bm(\mathbf x)}\sim \begin{cases}\frac{1}{\sqrt{N}}&\text{for $d=1$},\\
\sqrt{\frac{\ln N}{N}}&\text{for $d=2$},\\
\frac{1}{\sqrt[d]{N}}&\text{for $d\geq 3$}.
\end{cases}\label{scaling}
\end{equation}
In the present paper we are interested in the correlation function of the optimal matching ray $\Bm$ in the large $N$ limit; we will assume periodic boundary conditions on the hypercube $\Omega_d$, i.e., we will consider the problem on the flat hypertorus $\mathsf T^d$ in $d$ dimensions. We will analyze both the case in which two sets of random points are considered and the case in which one set of points is supposed fixed on a regular hypercubic lattice, whilst the second set is obtained from a Poisson process (see, e.g., fig.~\ref{toro} for a pictorial representation of a realization of the two dimensional problem). To introduce our results, in Section \ref{MK} we will review the Monge--Kantorovi\v{c} formulation of the optimal transport problem, from which a suitable general ansatz, already used by \textcite{Caracciolo2014}, is derived for the expression of the optimal matching ray in the continuum limit for the $p=2$ case. Using this working ansatz, in Section \ref{CF}, we will consider the \textsc{Ebmp} with quadratic cost on $\mathsf T^d$, and we will give evidences that, in the large $N$ limit, the correlation function
\begin{equation}
C_d(\mathbf x)=\left.\overline{\Bm(\mathbf r_i)\cdot\Bm(\mathbf r_j)}\right|_{\mathbf r_i-\mathbf r_j=\mathbf x}
\end{equation}
is related to the Green's function of the Laplacian operator on $\mathsf T^d$. We will consider also the correlation function for the normalized optimal matching ray
\begin{equation}
\boldsymbol\sigma(\Br_i)\coloneqq\frac{\Bm(\Br_{i})}{\norm{\Bm(\Br_{i})}}.
\end{equation}
Finally, we will give numerical evidences that the functional forms of the correlation functions obtained for $p=2$ in the two dimensional case are in good agreement with the numerical results for $p=1$ and $p=3$ in the same dimension.

To our knowledge, these results are new to the literature, where only the $d=1$ case is evaluated explicitly \citep{Boniolo2012,Caracciolo2014b}. The present work can be seen as a natural expansion and completion of a previous work of \textcite{Caracciolo2014}.

\section{The Monge--Kantorovi\v{c} mass transfer problem and correlation functions}\label{MK}

\subsection{The Monge--Kantorovi\v{c} problem}
The \textit{Monge--Kantorovi\v{c} transfer problem} is a well studied problem in Measure Theory \cite{Evans1997}, in the context of Transportation theory. Let us suppose that we are given two nonnegative smooth density measures $\rho_1$ and $\rho_2$ on a compact subset $\Omega\subset\R^d$, $\int_\Omega\rho_1(x)\dd ^dx=\int_\Omega\rho_2(x)\dd^d x$. Suppose also that a work function $w\colon \Omega\times \Omega\to\R^+$ is given. We want to find a one-to-one mapping $\BM\colon \Omega\to \Omega$, called \textit{optimal transport map}, such that the following conditions are satisfied:
\begin{enumerate}
\item $\BM\in\mathcal T$, $\mathcal T$ set of suitable \textit{transport maps}, defined as follows
\begin{multline}\textstyle
\mathcal T\coloneqq\left\{\mM\colon\Omega\to\Omega|\int_A\rho_1(\mathbf x)\dd^d x=\int_{\mM^{-1}(A)}\rho_2(\mathbf x)\dd^d x\right.\\\textstyle\left.\forall A\subseteq\Omega\text{ measurable}\right\}.\end{multline}
\item $\BM$ minimizes a certain cost functional
\begin{equation}\textstyle
E[\mM;w]\coloneqq\int_{\Omega} w(\mathbf x,\mM(\mathbf x))\rho_1(\mathbf x)\dd^d x,
\end{equation}
where $w\colon\Omega\times\Omega\to\R^+$ is a transportation cost function, i.e., $E[\BM;w]=\min_{\mM\in\mathcal T}E[\mM;w]$.
\end{enumerate}
Interestingly, it can be proven that, if $w(\mathbf x,\mathbf y)=\norm{\mathbf x-\mathbf y}^p$ with $p\in(1,+\infty)$ the condition $\mM\in\mathcal T$ can be re-expressed as a change-of-variable formula \cite{villani2008optimal}
\begin{equation}\label{cov}
\rho_1(\mathbf x)=\rho_2(\mM(\mathbf x))\det{\J_{\mM}(\mathbf x)},\quad \forall\mathbf x\in\Omega,
\end{equation}
where $\J_{\mM}(\mathbf x)$ is the Jacobian matrix of $\mM$, $\left(\J_{\mM}(\mathbf x)\right)_{ij}\equiv\frac{\partial\mathcal M_i}{\partial x_j}(\mathbf x)$. Moreover, if $p=2$, it can be proved \citep{villani2008optimal} that the optimal transport map can be expressed as a gradient of a scalar potential $\varphi$, i.e.,
\begin{equation}\label{vf}
\BM(\mathbf x)=\nabla\varphi(\mathbf x).
\end{equation}
It follows that the solution $\BM$ of the transport problem has to be identified among the vector fields in the form \eqref{vf}, with $\varphi$ satisfying the following relation:
\begin{equation}\label{MA}
\rho_1(\mathbf x)=\rho_2(\nabla\varphi(\mathbf x))\det{\Hess\varphi(\mathbf x)},\quad \forall\mathbf x\in\Omega.
\end{equation}
In the previous expression $\Hess\varphi(\mathbf x)$ is the Hessian matrix, $\left(\Hess\varphi(\mathbf x)\right)_{ij}=\frac{\partial^2\varphi}{\partial x_i\partial x_j}(\mathbf x)$. The nonlinear equation \eqref{MA} is known to the literature as \textit{Monge--Amp\`{e}re equation}.

Consider now the Monge--Kantorovi\v{c} problem in $\Omega\subset\R^d$, $\int_\Omega\rho_1(x)\dd^d x=\int_\Omega\rho_2(x)\dd^d x=1$, with work function $w(\mathbf x,\mathbf y)=\norm{\mathbf x-\mathbf y}^2$; suppose moreover that 
\begin{equation}
\rho_1(\mathbf x)=1+\delta\rho_1(\mathbf x)\text{ and }\rho_2(\mathbf x)=1+\delta\rho_2(\mathbf x),
\end{equation}
where
\begin{equation}
\abs{\delta\rho_1(\mathbf x)}\ll 1\text{ and }\abs{\delta\rho_2(\mathbf x)}\ll 1\quad \forall \mathbf x\in\Omega.
\end{equation}
 We expect that $\BM(\mathbf x)=\mathbf x+\Bm(\mathbf x)$, $\norm{\Bm(\mathbf x)}\ll 1$ $\forall \mathbf x\in\Omega$: in the first order approximation, $\det\J_{\BM}(\mathbf x)\approx 1+\nabla\cdot\Bm(\mathbf x)$, so we have that
\begin{equation}
\nabla\cdot\Bm(\mathbf x)\approx \rho_1(\mathbf x)-\rho_2(\mathbf x)\eqqcolon\rho(\mathbf x);
\end{equation}
in particular, using the fact that $\Bm=\nabla\phi$, we have that in the limit of our approximation the Poisson equation holds:
\begin{equation}
\Delta\phi(\mathbf x)=\rho(\mathbf x).\label{poiss}
\end{equation}
Note that in this case the total cost of the transport is given by \begin{equation}\textstyle E[\BM,\norm{\bullet}^2]=\int_{\Omega} \norm{\nabla\phi(\mathbf x)}^2\dd^d x.\end{equation} 

In the following we will consider the unit hypercube, $\Omega\equiv\Omega_d$, with periodic boundary conditions, i.e., we will work on the flat hypertorus $\mathsf{T}^d\coloneqq\R^d/\Z^d$. Being $\int_{\Omega_d}\rho(\mathbf x)\dd^d x=0$, Eq.~\eqref{poiss} has a unique solution on the compact manifold $\mathsf T^d$, given by
\begin{equation}\textstyle
\phi(\mathbf x)=\int_{\mathsf T^d} \rho(\mathbf y)G_d(\mathbf y,\mathbf x)\dd^d y,\label{solpoiss}
\end{equation}
where $G_d$ is the Green's function for the Laplace operator $\Delta$ on $\mathsf T^d$ defined by the relation 
\begin{equation}
\Delta_\mathbf{y} G_d(\mathbf x,\mathbf y)=\delta^{(d)}(\mathbf x-\mathbf y)-1,
\end{equation}
the solution of which can be written as
\begin{equation}
\label{green}G_d(\mathbf x,\mathbf y)\equiv G_d(\mathbf x-\mathbf y)=-\sum_{\mathbf n\in\Z^d\setminus\{\boldsymbol 0\}}\frac{\e^{2\pi i\mathbf n\cdot(\mathbf x-\mathbf y)}}{4\pi^2\norm{\mathbf n}^2}.
\end{equation}
In this linear approximation, the transport cost is given by
\begin{align}
E[\BM,\norm{\bullet}^2]&\textstyle=-\iint_{\mathsf T^d} \rho(\mathbf x)G_d(\mathbf x,\mathbf y)\rho(\mathbf y)\dd^d y\dd^d x\\&\textstyle=\sum_{\mathbf n\in\Z^d\setminus\{\boldsymbol 0\}}\frac{\abs{\hat\rho(\mathbf n)}^2}{4\pi^2\norm{\mathbf n}^2},\label{costocoulomb0}
\end{align}
where
\begin{equation}
\textstyle
\hat\rho(\mathbf n)\coloneqq\int_{\mathsf T^d}\rho(\mathbf x)\e^{-2\pi i\mathbf n\cdot\mathbf x}\dd^d x. 
\end{equation}
\subsection{Correlation functions for the \textsc{EBMP} on the hypertorus}\label{CF}
In the previous section we introduced the Monge--Kantorovi\v{c} transport problem and we obtained also a set of simple results for the transport problem with quadratic cost between two almost uniform measures, through a proper linearization. We want to extract useful information about the discrete combinatorial problem from the continuum problem using the fact that, in the large $N$ limit, the \textsc{Ebmp} between two sets of points on the hypertorus $\mathsf T^d$ appears as a transport problem between two atomic measures that can be assumed as almost uniform measures on the domain of interest. This na\"{i}ve approach is justified \textit{a posteriori} by the excellent agreement between theoretical predictions and numerical results. In particular, let us denote by $\mB\coloneqq\{\Bb_i\}_{i=1,\dots,N}\subset\Omega_d$ and $\mR\coloneqq\{\Br_i\}_{i=1,\dots,N}\subset\Omega_d$ two sets of points in $\Omega_d$, each set of cardinality $N$. The optimal cost of the matching on the flat hypertorus $\mathsf T^d$ with quadratic cost is therefore given, in the notation above, by
\begin{equation}\label{cost}E_{N}^{(2)}[\Bm;\{\mR,\mB\}]=\frac{1}{N}\sum_{i=1}^{N}\norm{\Bm(\Br_i)}^2.\end{equation}
In the previous formula, $\Bm$ is the (geodesic) optimal matching ray on $\mathsf T^d$. Let us now introduce two atomic measure densities:
\begin{subequations}\begin{eqnarray}
\rho_\mR(\mathbf x)&=\frac{1}{N}\sum_{i=1}^{N}\delta^{(d)}\left(\mathbf x-\Br_i\right),\\ \rho_\mB(\mathbf x)&=\frac{1}{N}\sum_{i=1}^{N}\delta^{(d)}\left(\mathbf x-\Bb_i\right).
\end{eqnarray}\end{subequations}
In the following, we will assume, as a working ansatz, that, in the large $N$ limit, $\Bm\to\Bm(\mathbf x)\equiv\nabla\phi(\mathbf x)$ in such a way that $\mathbf M(\mathbf x)=\mathbf x+\Bm(\mathbf x)$ is an optimal transport map between the two almost uniform measures above. Under the hypothesis that at least one set of points is randomly generated, we introduce the following correlation function
\begin{equation}
C_d(\mathbf x,\mathbf y)\coloneqq\overline{\nabla\phi(\mathbf x)\cdot\nabla\phi(\mathbf y)}
\end{equation}
and, by using Eq.~\eqref{solpoiss}, we obtain
\begin{multline}
C_d(\mathbf x,\mathbf y)\equiv C_d(\mathbf x-\mathbf y)=\\\textstyle=\iint \nabla_{\mathbf z}G_d(\mathbf z-\mathbf x)\cdot\nabla_{\mathbf w}G_d(\mathbf w-\mathbf y)\overline{\rho(\mathbf z)\rho(\mathbf w)}\dd^d z\dd^d w,
\label{corr}\end{multline}
where we denoted by 
\begin{equation}\begin{split}
\rho(\mathbf x)&\coloneqq\rho_\mR(\mathbf x)-\rho_\mB(\mathbf x)\\&=\frac{1}{N}\sum_{i=1}^{N}\left[\delta^{(d)}\left(\mathbf x-\Br_i\right)-\delta^{(d)}\left(\mathbf x-\Bb_i\right)\right]\end{split}
\end{equation}
and the average $\overline{\bullet}$ is intended over all possible instances. Observe now that $C_d(\mathbf 0)$ is, in the large $N$ limit, the average optimal cost for the Euclidean bipartite matching problem; using this simple correspondence, \textcite{Caracciolo2014} derived the correct scaling of the optimal cost and, through a proper regularization procedure, the finite size corrections to the average optimal cost for any dimension. In the following we will consider the complete correlation function $C_d(\mathbf x)$ in any dimension and we will derive it using the same ansatz successfully adopted by \textcite{Caracciolo2014} to obtain the scaling of the average optimal cost.  

We will distinguish two different cases.
\begin{description}
\item[Poisson--Poisson Euclidean matching problem] In the Poisson--Poisson (PP) Euclidean matching problem both the points of $\mR$ and the points of $\mB$ are random points uniformly distributed within $\Omega_d$. In this case we obtain
\begin{equation}
\overline{\rho(\mathbf x)\rho(\mathbf y)}=\frac{2}{N}\left[\delta^{(d)}(\mathbf x-\mathbf y)-1\right],
\end{equation}
and therefore the correlation function is
\begin{equation}\label{corrdpp}
C_d(\mathbf x-\mathbf y)=-\frac{2}{N}G_d(\mathbf x-\mathbf y).
\end{equation}
As anticipated, the average optimal cost is given by
\begin{equation}
E^{(2)}_N(d)\coloneqq\overline{E_{N}^{(2)}[\Bm;\{\mR,\mB\}]}=C_d(\mathbf 0).
\end{equation}
\item[Grid--Poisson Euclidean matching problem] In the grid--Poisson (\textsc{gP}) Euclidean matching problem  we suppose that $N=L^d$ for some natural number $L\in\N$ and that one set of points, e.g. the set $\mR=\{\Br_i\}_{i=1,\dots,N}$, is fixed on the vertices of an hypercubic lattice, in such a way that $\mR=\left\{\frac{\boldsymbol k}{L}|\boldsymbol k\in (0,L]^d\cap\N^d\right\}$, whilst the set $\mB=\{\Bb_i\}_{i=1,\dots,N}\subset\Omega_d$ is obtained as before considering randomly generated points in $\Omega_d$. We have
\begin{equation}\begin{split}
\overline{\rho(\mathbf x)\rho(\mathbf y)}=&\textstyle\frac{1}{N}\delta^{(d)}(\mathbf x-\mathbf y)+\frac{N^2-N}{N^2}\\&\textstyle+\frac{1}{N^2}\sum_{ij}\delta^{(d)}\left(\mathbf x-\Br_i\right)\delta^{(d)}\left(\mathbf y-\Br_j\right)\\&\textstyle-\frac{1}{N}\sum_{i}\left[\delta^{(d)}\left(\mathbf x-\Br_i\right)+\delta^{(d)}\left(\mathbf y-\Br_i\right)\right].
\end{split}\end{equation}
In this case the correlation function is therefore
\begin{equation}\label{corrdgp}
C^{\text{\textsc{gP}}}_d(\mathbf x-\mathbf y)=-\frac{1}{N}G_d(\mathbf x-\mathbf y).
\end{equation}
Being the average optimal cost of the matching in the grid-Poisson case
\begin{equation}
E_N^{(2;\textsc{gP})}(d)\coloneqq\overline{E_{N}^{(2)}[\Bm;\{\mR\text{ fixed},\mB\}]}=C^{\text{\textsc{gP}}}_d(\mathbf 0)
\end{equation}
we expect that in this case it will be asymptotically one half of the PP case.
\end{description}
We will consider also the correlation function for the normalized transport field, i.e. the following quantity:
\begin{equation}\label{corrfunnorm}
c_d(\mathbf x-\mathbf y)=\overline{\boldsymbol\sigma(\mathbf x)\cdot\boldsymbol\sigma(\mathbf y)},
\end{equation}
in which the correlation between the values normalized transport field \begin{equation}\boldsymbol\sigma(\mathbf x)\coloneqq\frac{\Bm(\mathbf x)}{\norm{\Bm(\mathbf x)}}=\frac{\nabla_\mathbf{x}\phi(\mathbf x)}{\norm{\nabla_\mathbf{x}\phi(\mathbf x)}}\end{equation}
in different positions is evaluated. Note that $\boldsymbol\sigma$ lives on the $d$-dimensional unit sphere.  To compute the correlation function \eqref{corrfunnorm} for the normalized field in the PP case, we assume a Gaussian behavior for the joint probability distribution of two values of the optimal transport field, and therefore we have
\begin{multline}
c_d(\mathbf x-\mathbf y)=\\=\iint\dd^d m_1\dd^d m_2\frac{\mathbf m_1\cdot\mathbf m_2}{\norm{\mathbf m_1}\norm{\mathbf m_2}}\frac{\e^{-\frac{1}{2}(\begin{smallmatrix}
\Bm_1&\Bm_2
\end{smallmatrix})\cdot\boldsymbol\Sigma^{-1}(\mathbf x,\mathbf y)\cdot\left(\begin{smallmatrix}
\Bm_1\\\Bm_2
\end{smallmatrix}\right)}}{\left(2\pi\sqrt{\det\boldsymbol\Sigma}\right)^d}
\end{multline}
where $\boldsymbol\Sigma(\mathbf x,\mathbf y)$ is the covariance matrix, 
\begin{align}
\boldsymbol\Sigma(\mathbf x,\mathbf y)&\coloneqq\begin{pmatrix}
\overline{\Bm(\mathbf x)\cdot \Bm(\mathbf x)}&&\overline{\Bm(\mathbf x)\cdot \Bm(\mathbf y)}
\\\overline{\Bm(\mathbf y)\cdot \Bm(\mathbf x)}&&\overline{\Bm(\mathbf y)\cdot \Bm(\mathbf y)}
\end{pmatrix}\\&\equiv\begin{pmatrix}
C_d(\mathbf 0)&&C_d(\mathbf x-\mathbf y)
\\C_d(\mathbf x-\mathbf y)&&C_d(\mathbf 0)
\end{pmatrix}
\end{align}
For $d\geq 2$ (the case $d=1$ was studied in \citep{Boniolo2012}), introducing
\begin{subequations}\begin{eqnarray}
A\coloneqq\frac{C_d(\mathbf 0)}{\det\boldsymbol\Sigma(\mathbf x,\mathbf y)},\\ B\coloneqq \frac{C_d(\mathbf x-\mathbf y)}{\det\boldsymbol\Sigma(\mathbf x,\mathbf y)},
\end{eqnarray}\end{subequations}
observe that $\frac{B}{A}\to 0^+$ for $N\to\infty$, being $NC_d(\mathbf x)$ finite for $\mathbf x\neq 0$ and $NC_d(\mathbf 0)\sim N^{1-\frac{2}{d}}$ for $d>2$, $NC_d(\mathbf 0)\sim \ln N$ for $d=2$. We have therefore that, in the notation above,
\begin{equation}
\det\boldsymbol\Sigma=\frac{1}{A^2-B^2}
\end{equation}
and
\begin{widetext}\begin{equation}\label{corrss}\begin{split}c_d(\mathbf x,\mathbf y)=&\textstyle\left(\frac{\sqrt{A^2-B^2}}{2\pi}\right)^d\frac{2\pi^\frac{d}{2}}{\Gamma\left(\frac{d}{2}\right)}\frac{2\pi^{\frac{d-1}{2}}}{\Gamma\left(\frac{d-1}{2}\right)}\int_0^{\pi}\dd\theta\sin^{d-2}\theta\cos\theta\int_0^\infty\dd m_1\int_0^\infty\dd m_2\, m_1^{d-1}m_2^{d-1}\e^{-\frac{A}{2}\left(m_1^2+m_2^2\right)+Bm_1m_2\cos\theta}\\=&\frac{B}{A}\frac{2\Gamma^2\left(\frac{d+1}{2}\right)}{d\Gamma^2\left(\frac{d}{2}\right)}\left(1-\frac{B^2}{A^2}\right)^\frac{d}{2}\pFq{2}{1}{\frac{d+1}{2}\,\frac{d+1}{2}}{\frac{d}{2}+1}{\frac{B^2}{A^2}}\xrightarrow[\frac{B}{A}\to 0]{N\to\infty}\frac{2}{d}\left(\frac{\Gamma\left(\frac{d+1}{2}\right)}{\Gamma\left(\frac{d}{2}\right)}\right)^2\frac{C_d(\mathbf x-\mathbf y)}{E_N^{(2)}(d)}.\end{split}\end{equation}\end{widetext}
In the previous expression, we have introduced the hypergeometric function
\begin{equation}
\pFq{2}{1}{a,b}{c}{z}\coloneqq\sum_{n=0}^\infty\frac{(a)_n(b)_n}{(c)_n}\frac{z^n}{n!},\quad (a)_n\coloneqq\frac{\Gamma\left(a+1\right)}{\Gamma\left(a-n+1\right)}.
\end{equation}
Observe that we can reproduce exactly the same calculation for the normalized field in the \textsc{gP}, obtaining
\begin{equation}\label{corrssgp}
c_d^{\textsc{gP}}(\mathbf x-\mathbf y)=\frac{2}{d}\left(\frac{\Gamma\left(\frac{d+1}{2}\right)}{\Gamma\left(\frac{d}{2}\right)}\right)^2\frac{C_d^{\textsc{gP}}(\mathbf x-\mathbf y)}{E_N^{(2;\textsc{gP})}(d)}.
\end{equation}
Finally, for $d\geq 2$ we can compute also the so called \textit{wall-to-wall correlation function} for the PP case:
\begin{equation}\label{wtw}\begin{split}
W_d(r)&\textstyle\coloneqq \prod_{i=2}^d\left(\int_0^1\dd x_i\right)c_d(r,x_2,\dots,x_d)\\
&\textstyle=-\frac{4}{dN}\left(\frac{\Gamma\left(\frac{d+1}{2}\right)}{\Gamma\left(\frac{d}{2}\right)}\right)^2\frac{G_1(r)}{E_N^{(2)}(d)}.\end{split}
\end{equation}
Similarly, the computation for the \textsc{gP} case gives
\begin{equation}\begin{split}
W^{\textsc{gP}}_d(r)&\textstyle\coloneqq \prod_{i=2}^d\left(\int_0^1\dd x_i\right)c^{\textsc{gP}}_d(r,x_2,\dots,x_d)\\
&\textstyle=-\frac{2}{dN}\left(\frac{\Gamma\left(\frac{d+1}{2}\right)}{\Gamma\left(\frac{d}{2}\right)}\right)^2\frac{G_1(r)}{E_N^{(2;\textsc{gP})}(d)}.\end{split}
\end{equation}

\section{Numerical results}
In the following we consider explicitly the cases $d=1$, $d=2$ and $d=3$ and we numerically verify the results presented above.
\subsection{Case $d=1$}

For $d=1$ we have that \begin{equation}G_1(r)=-\sum_{n\neq 0}\frac{1}{4\pi^2 n^2}\e^{2\pi i nr}=-\frac{1}{12}+\frac{\abs{r}}{2}\left(1-\abs{r}\right)\label{green1d}.\end{equation} It follows from Eq.~\eqref{corrdpp} that
\begin{equation}\textstyle
\label{corr1pp}
C_1(x-y)=\frac{1}{N}\left[\frac{1}{6}-\abs{x-y}\left(1-\abs{x-y}\right)\right];
\end{equation}
moreover, note that the average optimal cost is given by \begin{equation}\textstyle
\label{ene1pp}
E_N^{(2)}(1)=C_1(0)=\frac{1}{6N}.
\end{equation}

In the \textsc{gP} case we obtain from Eq.~\eqref{corrdgp}
\begin{equation}\textstyle
\label{corr1gp}
C^{\text{\textsc{gP}}}_1(x-y)=\frac{1}{N}\left[\frac{1}{12}-\abs{x-y}\frac{1-\abs{x-y}}{2}\right].
\end{equation}
The average total cost of the optimal matching is given by
\begin{equation}\textstyle
\label{ene1gp}
E_N^{(2;\textsc{gP})}(1)=C^{\text{\textsc{gP}}}_1(0)=\frac{1}{12N}.
\end{equation}
The previous results for the $d=1$ case are known to the literature \cite{Boniolo2012,Caracciolo2014,Caracciolo2014b}, although the correlation function was derived using a different probabilistic approach. In \cite{Boniolo2012} the correlation function for the normalized transport field is evaluated as
\begin{equation}\textstyle
c_1(x)=c_1^{\text{\textsc{gP}}}(x)=\frac{2}{\pi}\arctan\left[\frac{1-6x(1-x)}{\sqrt{12x(1-x)\left(1-3x(1-x)\right)}}\right].
\end{equation}

\subsection{Case $d=2$}
\begin{figure*}%
\subfloat[\label{g2figurenum}Section $C_2(r_1,0)$ and ${C^{\text{\textsc{gP}}}_2(r_1,0)}$ of the correlation function both in the PP case for $N=10^4$ and in the \textsc{gP} case for $N=3600$ and corresponding theoretical predictions.]{
\includegraphics[width=0.9\columnwidth]{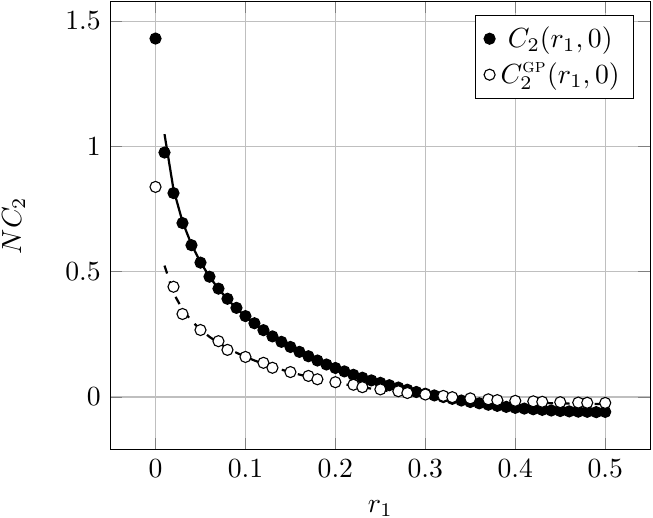}}\hspace{0.1\columnwidth}
\subfloat[\label{g2figurenumN} Section $c_2(r_1,0)$ for $N=10^4$ and ${c^{\text{\textsc{gP}}}_2(r_1,0)}$ for $N=3600$ of the correlation function and theoretical predictions, Eq.~\eqref{corrss} and Eq.~\eqref{corrssgp}: note that the theoretical curves overlap.]{\includegraphics[width=0.9\columnwidth]{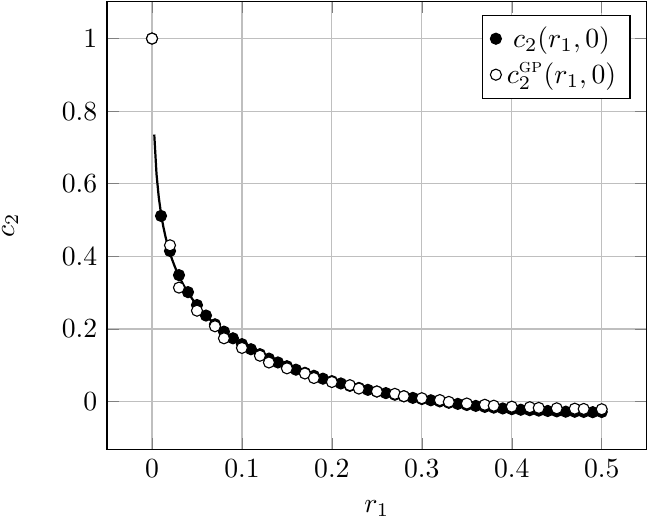}}\\
\subfloat[\label{wtwpp} Rescaled wall-to-wall correlation function in two dimensions for the PP matching problem with $N=3600$ on the unit flat torus. The continuous line corresponds to the analytical prediction.]{
\includegraphics[width=0.9\columnwidth]{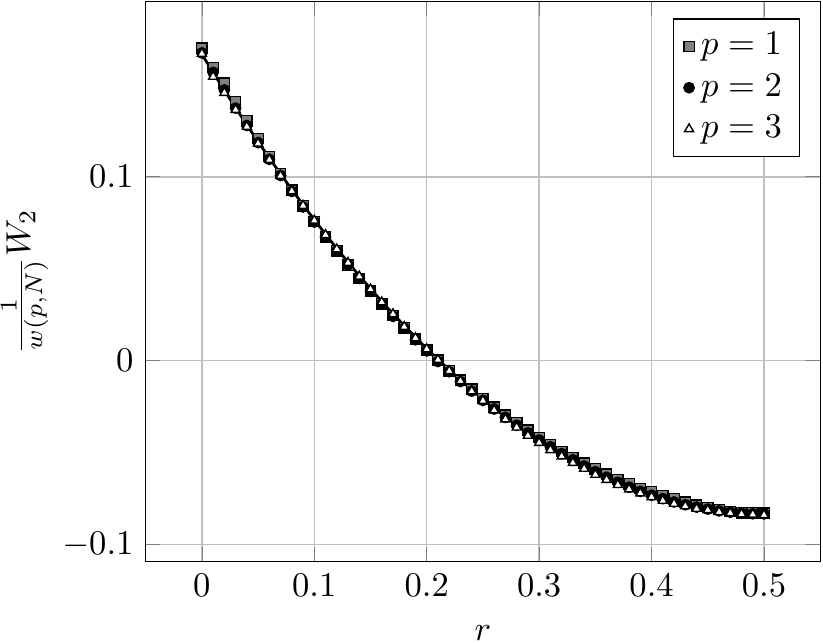}}\hspace{0.1\columnwidth}
\subfloat[\label{wtwgp} Rescaled wall-to-wall correlation function in two dimensions for the \textsc{gP} matching problem with $N=3600$ on the unit flat torus. The continuous line corresponds to the analytical prediction.]{
\includegraphics[width=0.9\columnwidth]{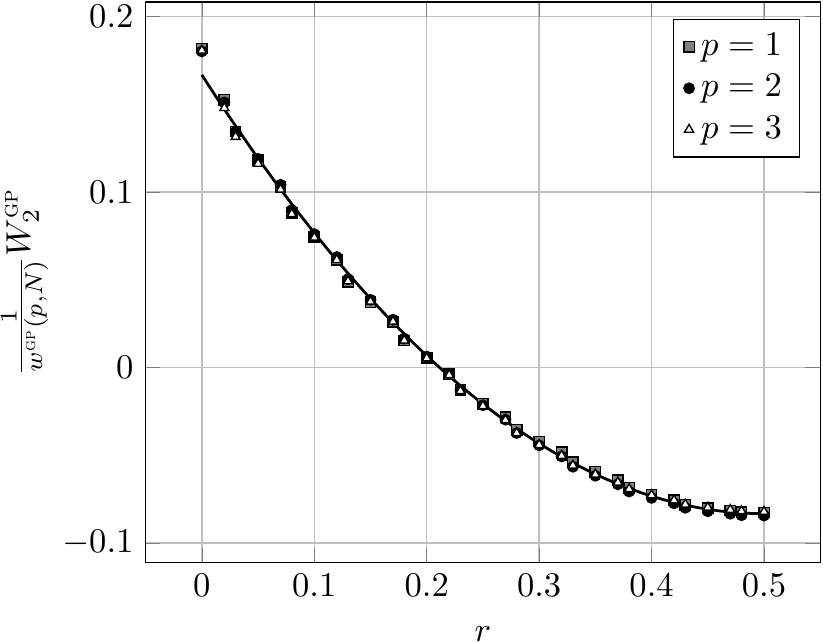}}
\caption{Theoretical predictions for the correlation functions $C_2(\mathbf x)$ and $c_2(\mathbf x)$ for the Euclidean bipartite matching problem in two dimensions and numerical results. Numerical results for the wall-to-wall correlation functions and corresponding theoretical predictions are also presented.}
\end{figure*}

For $d=2$, denoting by $\mathbf x=(x_1,x_2)$ and by $\mathbf y=(y_1,y_2)$ two points on the unit flat torus $\mathsf T^2$, the Laplacian Green's function can be written in terms of special functions as \cite{Lin2010}
\begin{multline}\textstyle
G_2(\mathbf x-\mathbf y)=-\sum_{\mathbf n\neq \mathbf 0}\frac{1}{4\pi^2 \norm{\mathbf n}^2}\e^{2\pi i \mathbf n\cdot(\mathbf x-\mathbf y)}\\\textstyle=\frac{1}{2\pi}\left.\ln\abs{2 \pi ^{3/4}\frac{\vartheta_1(\pi z|i)}{\Gamma \left(\frac{1}{4}\right)}}\right|_{z=(x_1-y_1)+i(x_2-y_2)}-\frac{\left(x_2-y_2\right)^2}{2},
\end{multline}
where we introduced the first Jacobi theta function
\begin{equation}
\vartheta_1(z|\tau)\coloneqq 2\e^\frac{i\pi\tau}{4}\sum_{n=0}^\infty(-1)^n \e^{i\pi\tau n(n+1)}\sin\left[(2n+1)z\right].
\end{equation}
From Eq.~\eqref{corrdpp} we have simply
\begin{equation}
\label{corr2pp}
C_2(\mathbf x)=-\frac{1}{N}\left[\frac{1}{2\pi}\left.\ln\abs{2 \pi ^{3/4}\frac{\vartheta_1(\pi z|i)}{\Gamma \left(\frac{1}{4}\right)}}\right|_{z=x_1+ix_2}-\frac{x_2^2}{2}\right].
\end{equation}
In the \textsc{gP} case, we have as usual
\begin{equation}
\label{corr2gp}
C^{\text{\textsc{gP}}}_2(\mathbf x-\mathbf y)=\frac{1}{2}C_2(\mathbf x-\mathbf y).
\end{equation}
Observe that the previous expressions contains no free parameters and therefore a direct comparison with numerical data is possible. We present our numerical results both for the \textsc{gP} case and the PP case in fig.~\ref{g2figurenum}. The average optimal cost for the PP \textsc{Ebmp} is given by $E_N^{(2)}(2)=C_2(\mathbf 0)$: however, $G_2(\mathbf x)$ is divergent for $\mathbf x=\mathbf 0$. Analysing the scaling of the total cost and performing a proper regularization of the previous quantity, \citet{Caracciolo2014} obtained:
\begin{equation}\label{e2pp}
C_2(\mathbf 0)=\frac{1}{N}\left(\frac{\ln N}{2\pi}+\beta_\text{\textsc{pp}}\right)+o\left(\frac{1}{N}\right),\quad\beta_\text{\textsc{pp}}=0.1332(5).
\end{equation} 
A numerical fit of the optimal costs for $d=2$ for the \textsc{gP} \textsc{Ebmp} gives
\begin{equation}\label{e2gp}
C_2^\text{\textsc{gP}}(\mathbf 0)=\frac{1}{2N}\left(\frac{\ln N}{2\pi}+\beta_\text{\textsc{gP}}\right)+o\left(\frac{1}{N}\right),\ \beta_\text{\textsc{gP}}=0.3758(5).
\end{equation}

The correlation function \eqref{corrfunnorm} for the normalized matching field in the PP case has the expression \eqref{corrss}, 
\begin{equation}
c_2(\mathbf x-\mathbf y)=\frac{\pi}{4}\frac{C_2(\mathbf x-\mathbf y)}{C_2(\mathbf 0)}.
\end{equation}
Observe that the only free parameter in this quantity is $C_2(\mathbf 0)$: inserting the value obtained by \textcite{Caracciolo2014}, Eq.~\eqref{e2pp}, we obtain the theoretical prediction in fig.~\ref{g2figurenumN}, where we also present some numerical results for $c_2(\mathbf x,\mathbf y)$ that show the agreement with the theoretical curve. 

Using the value \eqref{e2gp} in Eq.~\eqref{corrssgp} for $d=2$ we also obtained the theoretical curve for the grid--Poisson problem depicted in fig.~\eqref{g2figurenumN}, where, once again, an excellent agreement is found with numerical data.

Finally, let us compute the wall-to-wall correlation function for the PP Euclidean matching problem on the flat torus. The theoretical prediction is given by Eq.~\eqref{wtw},
\begin{equation}\label{wtweq}
W_2(r)=-\frac{\pi}{2NC_2(\mathbf 0)}G_1(r).
\end{equation}
In the \textsc{gP} case, instead, we have
\begin{equation}\label{wtweqgp}
W_2^{\text{\textsc{gP}}}(r)=-\frac{\pi}{4 NC^{\text{\textsc{gP}}}_2(\mathbf 0)}G_1(r).
\end{equation}
Numerical results both for the PP case and for the \textsc{gP} case are presented in fig.~\ref{wtwpp} and fig.~\ref{wtwgp}. Once again observe that the values of the average optimal cost in the corresponding cases, Eq.~\eqref{e2pp} and Eq.~\eqref{e2pp}, fix completely the expression of the wall-to-wall correlation function.

\subsubsection*{The case $p\neq 2$}

\begin{table}
\begin{tabular}{cc|cc}
\toprule
&&$\alpha(p)$&$\beta$ fixed\\
\colrule
\multirow{3}{*}{PP}&$p=1$&$0.860(5)$&\multirow{3}{*}{$\beta\equiv \beta_\text{\textsc{pp}}=0.1332(5)$}\\
&$p=2$&$0.996(5)$&\\
&$p=3$&$0.96(1)$&\\\colrule
\multirow{3}{*}{\textsc{gP}}&$p=1$&$0.88(1)$&\multirow{3}{*}{$\beta\equiv \beta_\text{\textsc{gP}}=0.3758(5)$}\\
&$p=2$&$1.02(3)$&\\
&$p=3$&$0.98(2)$&\\
\botrule
\end{tabular}
\caption{Fit results for the wall-to-wall correlation function of the Euclidean matching problem for $d=2$ and $p=1,2,3$ with reference to the notation of the fitting curve, Eq. \eqref{wtwgenerale}. Observe that we expected $\alpha(2)=1$.}\label{wtwtable}
\end{table}

Up to now, we analyzed the correlation functions for $p=2$ and $d=2$ obtained from the linearized equation \eqref{poiss}. We present some numerical results for $p=1$ and $p=3$ and $N=3600$ both in the \textsc{gP} case and in the PP in the two dimensional case. In particular, we analyzed the wall-to-wall correlation function for different values of $p$ and we obtained numerical evidences of a functional form for it of the type
\begin{equation}\textstyle
W_2(r;p,N)=w(p,N)\left[\frac{1}{6}-r(1-r)\right]\label{wtwgenerale}
\end{equation}
both in the PP and in the \textsc{gP} case. Inspired by the obtained expression for the $p=2$ case, Eq.~\eqref{wtweq}, we assumed for the global factor $w(p)$ the following dependence on the size $N$ of the considered system
\begin{equation}
w(p,N)=\frac{\alpha(p)\pi^2}{2\ln N+4\pi\beta}.
\end{equation}
where $\alpha(p)$ depends only on the weight exponent $p$ and $\beta\equiv\beta_\text{\textsc{gP}}$ if we are considering a \textsc{gP} matching, $\beta\equiv \beta_\text{\textsc{pp}}$ if we are dealing with a PP matching. We expected that $\alpha(2)=1$. We performed a numerical fit using the previous expression also for $p=2$, obtaining the results presented in Table~\ref{wtwtable}. Numerical data are in excellent agreement with the functional expression \eqref{wtwgenerale}, suggesting therefore that the wall-to-wall correlation function in $d=2$ has the same expression for all values of $p$ up to a non universal multiplicative constant depending on the exponent that appears in the weight function. However, further investigations in this direction are needed to confirm this results in a wider range of values of $p$.

\subsection{Case $d=3$}
\begin{figure}
\includegraphics[width=\columnwidth]{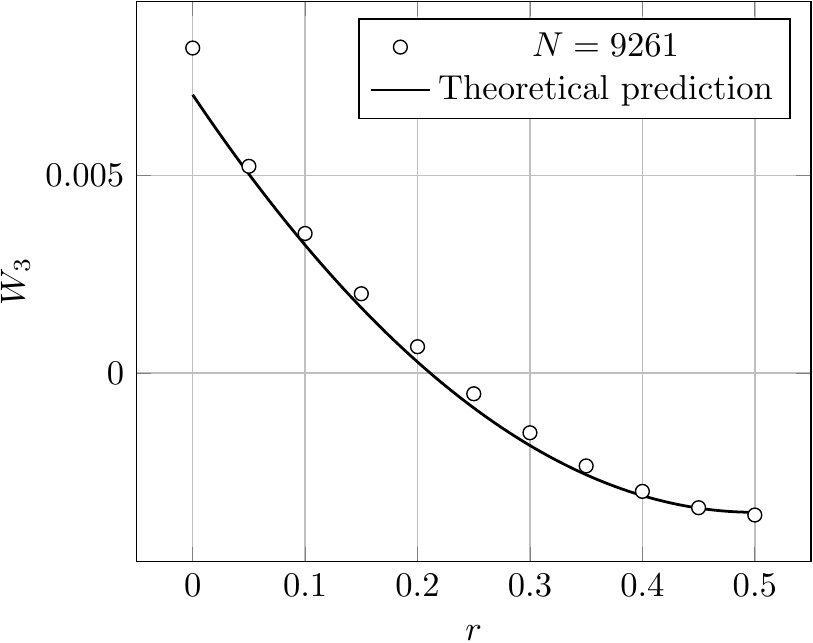}
\caption{Wall-to-wall correlation function in three dimensions for the \textsc{gP} matching problem with $d=3$ and $N=9261$ on the unit flat hypertorus.
}\label{wtwplot3dpp}
\end{figure}
The general expressions, Eq.~\eqref{corrdpp} and Eq.~\eqref{corrdgp}, for the correlation functions presented above can be applied in any dimensionality in the $N\to\infty$ limit. We show here some numerical results for $d=3$ for the grid--Poisson case, taking however into account that the precision of these data is lower, since the computation has complexity $O(N^3)$ (i.e., the computational complexity of the Hungarian algorithm in the \textcite{Edmonds1972} version), where the number of points $N$ has to scale exponentially with the system dimensionality in order to maintain the same accuracy.

For $d=3$ Eq.~\eqref{green} and Eq.~\eqref{corrdpp} give
\begin{equation}
C_3(\mathbf x-\mathbf y)=\frac{1}{2\pi^2N}\sum_{\mathbf n\in\Z^3\setminus\{\mathbf 0\}}\frac{1}{\norm{\mathbf n}^2}\e^{2\pi i\mathbf n\cdot(\mathbf x-\mathbf y)}.
\end{equation}
Clearly the previous function can not be represented in a plot. From the correlation function $C_3(\mathbf x)$, the wall to wall correlation function can be obtained as before in the form \begin{equation}\textstyle W_3(r)=-\frac{16}{3\pi NC_3(\mathbf 0)}G_1(\mathbf r).\end{equation} As in the previous cases, $C_3(\mathbf 0)$ can be evaluated from the cost fit \cite{Caracciolo2014} and it is equal to $C_3(\mathbf 0)=0.66251(2)N^{-\frac{2}{3}}-\frac{0.45157\dots}{N}$ (note that an exact formula for the coefficient of the $\frac{1}{N}$ correction to $C_3(\mathbf 0)$ is provided in \cite{Caracciolo2014} in terms of an Epstein function).

Following the same procedure of the PP case, we can compute the wall-to-wall correlation function on the unit hypercube in $d=3$ for the \textsc{gP} matching problem. Reproducing the computations of the $d=2$ case we have 
\begin{equation}\label{wtw3d}\textstyle
W_3^{\text{\textsc{gP}}}(r)=-\frac{8}{3\pi NC^{\text{\textsc{gP}}}_3(\mathbf 0)}G_1(r).
\end{equation}
We evaluated $C_3^{\text{\textsc{gP}}}(\mathbf 0)$ from the cost scaling, obtaining
\begin{equation}\textstyle
C_3^{\text{\textsc{gP}}}(\mathbf 0)=0.4893(4) N^{-\frac{2}{3}}-\frac{0.23(5)}{N}.
\end{equation}
The prediction obtained and the numerical data are presented in fig.~\ref{wtwplot3dpp}.

\section{Conclusions}
In the present work we adopted the scaling ansatz proposed by \textcite{Caracciolo2014} to compute the correlation function for the optimal matching ray and for the normalized optimal matching ray in the Euclidean bipartite matching problem on the $d$-dimensional flat hypertorus with quadratic cost. We showed also that the correlation function is strictly related to the Green's function of the Laplacian operator on the flat hypertorus itself in the large $N$ limit. Given the value the average optimal cost at fixed size $N$, the obtained expressions have no free parameters and were directly compared with the results of numerical simulations, showing an excellent agreement. For $d=2$ and $d+3$ we computed also the wall-to-wall correlation function: for $d=2$ in particular we give numerical evidences that, for $p\neq 2$, the wall-to-wall correlation function has the same form obtained for $p=2$ up to a global multiplicative constant. 

All previous results suggest that the \textsc{Ebmp} with quadratic cost, in the large $N$ limit, appears as a Gaussian free theory on the $d$-dimensional flat hypertorus, in such a way that the correlation function of the matching ray is related directly to the free propagator of the theory itself. In subsequent publications we will investigate this crucial aspect of the problem and its implications on the universal behavior for different values of the exponent $p$ in the cost functional. 

\section{Acknowledgments}
We thank Luigi Ambrosio, from Scuola Normale Superiore di Pisa, Carlo Lucibello, from Politecnico di Torino, and Giorgio Parisi, from University of Rome ``La Sapienza'', for useful discussions. G.S. acknowledges partial financial support from the John Templeton Foundation.

\bibliography{Biblio.bib}
\end{document}